\documentclass[a4paper,11pt]{article}
\usepackage{amsmath, amssymb, amsfonts,indentfirst,graphicx,color,xcolor,anysize}
\usepackage{placeins}
\usepackage{hyperref}
\usepackage{xspace}
\usepackage{booktabs}

\usepackage[comma,square,numbers,sort&compress]{natbib}
\usepackage[a4paper, top=2cm, bottom=2cm, left=2cm, right=2cm]{geometry}
\usepackage{hyperref}
\usepackage{blindtext}

\marginsize{3cm}{3cm}{2cm}{2cm}
\title{
	\includegraphics[width=0.35\textwidth]{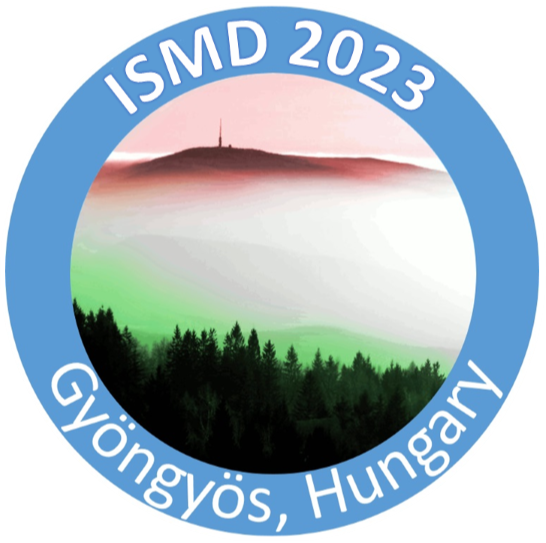}\\[1cm]
	\textbf{Non-prompt $\mathrm{J}/\psi$ production in proton-proton collisions with ALICE}}
\author{{Wenda Guo$^{1,2}$}\\(on behalf of the ALICE Collaboration)\\[1ex]
	$^1$Central China Normal University, Wuhan, China\\
	$^2$University of Bergen, Bergen, Norway\\
}
\begin{document}

\maketitle

\begin{abstract}
    ${\rm J/\psi}$ production in high-energy hadronic collisions is sensitive to both perturbative and non-perturbative aspects of quantum chromodynamics (QCD) calculations. The production of a heavy-quark pair is well-described by perturbative QCD, whereas, the
    formation of the bound state involves non-perturbative processes, treated in different ways by various available theoretical models. ALICE can measure  inclusive ${\rm J/\psi}$ at both forward and midrapidity down to zero $p_{\rm T}$, and the prompt and non-prompt ${\rm J/\psi}$ separation was performed at midrapidity in Run 1 and 2. The study of the production of non-prompt ${\rm J/\psi}$ originating from the decay of beauty hadrons, besides allowing the isolation the prompt ${\rm J/\psi}$ cross section from the inclusive ${\rm J/\psi}$ cross section, can be used to estimate open beauty-hadron production. In addition, heavy-flavour particle production in pp collisions as a function of charged-particle multiplicity can provide insight into processes occuring in the collisions at the partonic level, as well as the interplay between the hard and soft mechanisms in particle production.\\
\end{abstract}

\section{Introduction}
\label{sec:introduction}

Quarkonium production in hadronic interactions is an excellent case of study for understanding hadronization mechanisms in quantum chromodynamics (QCD), the theory of strong interactions \cite{Brambilla:2010cs}. Experimentally, the reconstruction of the $\mathrm{J}/\psi$ meson, produced in pp collisions at the energies of the Large Hadron Collider (LHC), gives access to both the physics of charmonium systems and that of beauty-quark production. 
Indeed, direct $\mathrm{J}/\psi$ mesons and feed-down from higher mass charmonium states such as $\chi_{\rm c}$ and $\psi(\rm 2S)$, which are denoted as the "prompt" component, can be experimentally separated from the contribution from long-lived weak decays of beauty hadrons, denoted as the "non-prompt" component. In addition, due to the large rest mass of the $\mathrm{J}/\psi$ as compared to other beauty-hadron decay products, the $\mathrm{J}/\psi$ momentum vector is very close to the one of the decaying beauty-hadron, making the non-prompt $\mathrm{J}/\psi$ measurement a good tool to study the production of beauty-flavour hadrons~\cite{PhysRevD.71.032001}. 

The inclusive production of open heavy-flavour hadrons in hadronic collisions is computed using the collinear factorisation approach~\cite{Collins:1989gx} as a convolution
of the parton distribution functions of the incoming hadrons, the hard parton-parton scattering cross section computed perturbatively, and the fragmentation process describing the non-perturbative evolution of a charm- or beauty-quark into an open heavy-flavour hadron.
These calculations are implemented at the next-to-leading order (NLO) accuracy in the general-mass variable-flavour-number scheme
(GM-VFNS)~\cite{Benzke:2019usl,Kniehl:2005mk}, and at NLO with an all-order resummation to next-to-leading log (NLL) accuracy in the limit where the $p_{\rm T}$ of the heavy quark is much larger than its mass in the FONLL resummation approach~\cite{Cacciari_2001,Cacciari:2012ny}.
Recent calculations with next-to-next-to-leading-order (NNLO) QCD radiative corrections are implemented for the beauty-quark production cross section~\cite{Catani:2020kkl}.
Other predictions are also performed in the leading order (LO) approximation through the $k_{\rm T}$-factorisation framework~\cite{Catani:1990eg}.
Non-prompt $\mathrm{J}/\psi$ production is directly related to open beauty-hadron production, and can be used to estimate the latter after an extrapolation.
The measurements of the total beauty-quark production cross sections  are less sensitive to the non-perturbative hadronisation effects than the total charm-quark production, which makes them a good test for QCD in the perturbative regime.

The event-multiplicity-dependent production of charmonium and open charm hadrons in pp and p--Pb collisions are observables with the potential to give new insights on processes at the parton level and into the interplay between the hard and soft mechanisms in particle production.
ALICE has studied the multiplicity dependence in pp collisions at $\sqrt{s}$ = 5.02, 7 and 13 TeV of inclusive $\mathrm{J}/\psi$ production at central~\cite{Abelev:2012rz, ALICE:Sweber} and forward rapidity~\cite{forwardmult13tev} of non-prompt $\mathrm{J}/\psi$ and D-meson production at midrapidity~\cite{Adam:2015ota}. The general observation is the increasing open and hidden charm production with the charged-particle multiplicity measured at midrapidity. Multiplicities of about 4 times the mean multiplicity value were reached at $\sqrt{s}$ = 7 TeV, and 7 times at $\sqrt{s}$ = 13 TeV for inclusive $\mathrm{J}/\psi$ production. The results are consistent with an approximately linear increase of the normalized yield at forward rapidity, and a faster-than-linear increase at midrapidity, as a function of the normalized multiplicity (both observables are normalized to their corresponding averages in minimum bias events). For the D-meson production at midrapidity, normalized event multiplicities of about 6 were reached, and a stronger-than-linear increase in the D-meson production was observed at the highest multiplicities.
Observations made by the CMS Collaboration for $\Upsilon(\mathrm{nS})$ production at midrapidity and $\sqrt{s}$ = 2.76 TeV indicate a linear increase with the event activity when measuring the latter at forward rapidity and a stronger-than-linear increase with the event activity measured at midrapidity~\cite{Chatrchyan:2013nza}.
At RHIC, the measurement of $\mathrm{J}/\psi$ production as a function of multiplicity was recently performed by the STAR Collaboration~\cite{Adam:2018jmp} for $\sqrt{s}$ = 200 GeV, showing similar trends as observed in the LHC data.
The $\mathrm{J}/\psi$ production as a function of charged-particle multiplicity was also studied in p--Pb collisions, exhibiting significant differences for different ranges of the rapidity of the $\mathrm{J}/\psi$ meson~\cite{Adamova:2017uhu,Acharya:2020giw}.
A clear correlation with the event multiplicity (and event shape) was experimentally established for the inclusive charged-particle production~\cite{Acharya:2019mzb} as well as for identified particles, including multi-strange hyperons~\cite{Acharya:2018orn}.

This proceeding will briefly introduce the ALICE setup dedicated to the reconstruction of event properties and tracks, in addition to measurements of  charmonium. Some published results of prompt and non-prompt $\mathrm{J}/\psi$ production cross sections~\cite{ALICE:jpsi:02, ALICE:jpsi:04} will be presented. Inclusive $\mathrm{J}/\psi$ production as a function of multiplicity in proton-proton collisions at $\sqrt{s}$ = 13 TeV~\cite{ALICE:Sweber}, as well as the non-prompt $\mathrm{J}/\psi$ production as a function of multiplicity in proton-proton collisions at $\sqrt{s}$ = 7 TeV~\cite{ALICE:jpsi:01}, will be shown. Besides, the likelihood fit technique used to separate prompt and non-prompt $\mathrm{J}/\psi$ and a series of performance results in proton-proton collisions at $\sqrt{s}$ = 13 TeV will be shown.

\section{ALICE detector in Run 2 and data samples}
\label{sec:detector}
A detailed description of the ALICE apparatus and its performance can be found in Refs.~\cite{Collaboration_2008,Abelev:2014ffa}. The main detectors of the central barrel employed for the reconstruction of the $\mathrm{J}/\psi$ via the ${\rm e^{+}e^{-}}$ decay channel are the Inner Tracking System (ITS)~\cite{Aamodt:2010aa} and the Time Projection Chamber (TPC)~\cite{ALME2010316}. Both are used for track reconstruction, while, in addition, the TPC is also used for electron identification and the ITS for primary and secondary vertex reconstruction. 
%The events are selected using a minimum bias trigger (MB) provided by the V0 detectors~\cite{Abbas:2013taa}, defined as the coincidence of a signal in the V0A and V0C subsystems, which are scintillator arrays placed on both sides of the nominal interaction point. The High-multiplicity trigger (HM) requires a signal amplitude in the V0 arrays above a threshold which corresponds to the 0.1\% highest multiplicity events.
Besides, the V0 detector, which consists of two forward scintillator arrays~\cite{Abbas:2013taa} covering the pseudorapidity ranges $-3.7<\eta<-1.7$ and $2.8<\eta<5.1$, provides minimum bias (MB) trigger and high multiplicity (HM) trigger. The MB trigger signal consists of a coincident signal in both arrays, while the HM trigger requires a signal amplitude in the V0 arrays above a threshold, corresponding to the 0.1\% highest multiplicity events.

The measurements are based on data samples collected during the years of 2016$-$2018 in pp collisions at $\sqrt{s}$ = 13 TeV. The event sample, which is the same as the one used for the published inclusive $\mathrm{J}/\psi$ analyses at the same energy~\cite{Alice13TeV}, corresponds to the integrated luminosity of $\mathcal{L}_{\rm int}$ = 32.2 $\pm$ 0.5 nb$^{-1}$~\cite{ALICE-PUBLIC-2021-005}. Besides, the samples used for the measurement at $\sqrt{s}~=~7~\mathrm{TeV}$~\cite{ALICE:jpsi:nonprompt:mult:7tev} were recorded in 2010 with the corresponding integrated luminosity of $\mathcal{L}_{\rm int}$ $\simeq$ 5 nb$^{-1}$. 

\section{Data analysis}
\label{Sec:dataAnalysis}

The strategies of the measurements at $\sqrt{s}$ =7 TeV and 13 TeV are similar, thus, the details of the analysis $\sqrt{s}~=~13~\mathrm{TeV}$ are shown in this section, while the procedure at $\sqrt{s}~=~7~\mathrm{TeV}$ can be found in Ref.~\cite{ALICE:jpsi:nonprompt:mult:7tev}.
Event selection and track quality requirements used in these analyses are identical to those used for the corresponding inclusive $\mathrm{J}/\psi$ cross section analyses at $\sqrt{s}$ = 13 TeV~\cite{Alice13TeV}.  Events are binned in multiplicity classes based on the signal on the Silicon Pixel Detector (SPD), which is the innermost layer of ITS. For the measurement of the charged-particle pseudorapidity density $\mathrm{d}N_{\rm ch}/\mathrm{d}\eta$ at midrapidity, $|\eta|<1$, the SPD tracklets, track segments joining hits in the two SPD layers~\cite{2023137782}, are used~\cite{Adam:2015pza}.

For the analysis using an event selection based on the forward multiplicity measurement with the V0 detector, the signal amplitudes are equalized to compensate for detector ageing and the small acceptance variation with the event vertex position.

The normalized charged-particle pseudorapidity density in each event class is calculated as~\cite{ALICE:Sweber}:

\begin{equation}
\frac{\mathrm{d}N_{\rm ch}/\mathrm{d}\eta}{\langle \mathrm{d}N_{\rm ch}/\mathrm{d}\eta \rangle_{\mathrm{INEL}>0}} = \frac{\beta \times \langle N^{\rm corr}_{\rm trk}\rangle}{\Delta \eta \times \langle \mathrm{d}N_{\rm ch}/\mathrm{d}\eta \rangle_{\mathrm{INEL}>0}},
\end{equation}

where $\langle N^{\rm corr}_{\rm trk}\rangle$ is the averaged value of $N^{\rm corr}_{\rm trk}$, in each multiplicity class, is corrected for the trigger and vertex finding efficiencies. $\beta$ is the correction factor to obtain the average $\mathrm{d}N_{\rm ch}/\mathrm{d}\eta$ value corresponding to a given $N^{\rm corr}_{\rm trk}$ interval via the $N^{\rm corr}_{\rm trk}-N_{\rm ch}$ correction introduced in Ref.~\cite{ALICE:jpsi:03}. The former is estimated from Monte Carlo (MC) simulations and the latter with a data-driven approach. Those efficiencies are below unity only for the low-multiplicity events.

To perform the charmonium signal extraction, selected tracks must have a minimum transverse momentum of 1 GeV/$c$ and a pseudorapidity in the range $|\eta| < 0.9$. 
The $\mathrm{J}/\psi$ candidates are formed by considering all opposite charge electron pairs.
Prompt $\mathrm{J}/\psi$ mesons are separated from those originating from beauty-hadron decays on a statistical basis, exploiting the displacement between the primary event vertex and the decay vertex of the $\mathrm{J}/\psi$. The measurement of the fraction of $\mathrm{J}/\psi$ mesons originating from beauty-hadron decays, $f_{\rm B}$, is carried out through an unbinned two-dimensional likelihood fit procedure, following the same technique adopted in the previous pp analysis~\cite{Abelev:2012gx}. A simultaneous fit of the dielectron pair invariant mass ($m_{\rm ee}$) and pseudoproper decay length ($x$) distributions is performed. The latter is defined as $x = c \times \Vec{L} \times \Vec{p_{\rm T}} \times m_{\mathrm{J}/\psi}/|\Vec{p_{\rm T}}|$, where $\Vec{L}$ is the vector pointing from the primary vertex to the $\mathrm{J}/\psi$ decay vertex and $m_{\mathrm{J}/\psi}$ is the $\mathrm{J}/\psi$ mass provided by the Particle Data Group (PDG)~\cite{Tanabashi:2018oca}. The fit procedure maximises the logarithm of a likelihood function:

\begin{equation}
\label{eq:likeFunc} 
\ln{\mathcal{L}}= \sum_{i=1}^{N} \ln{\left[ f_{\rm Sig} \times F_{\rm Sig}(x^{i}) \times M_{\rm Sig}(m_{\rm ee}^{i}) + (1 - f_{\rm Sig}) \times F_{\rm Bkg}(x^{i}) \times M_{\rm Bkg}(m_{\rm ee}^{i})\right]},
\end{equation}

where $N$ is the number of $\mathrm{J}/\psi$ candidates within the invariant-mass interval $2.4 < m_{\rm ee} < 3.6~\mathrm{GeV}/c^{2}$, $F_{\rm Sig}(x)$ and $F_{\rm Bkg}(x)$ represent the probability density functions (PDFs) for the pseudoproper decay length distributions of signal and background, respectively. Similarly, $M_{Sig}(m_{\rm ee})$ and $M_{\rm Bkg}(m_{\rm ee})$ represent the equivalent PDFs for the invariant mass distributions. The signal fraction within the invariant-mass window considered for the fit, $f_{\rm Sig}$, represents the relative fraction of signal candidates, both prompt and non-prompt, over the sum of signal and background. The pseudoproper decay length PDF of the signal is defined as: 

\begin{equation}
\label{eq:signalPseudoPropDL} 
F_{\rm Sig}(x) =  f^{'}_{\rm B} \times F_{\rm B}(x) + (1-f^{'}_{\rm B}) \times F_{\rm prompt}(x),
\end{equation}

where $F_{\rm B}(x)$ and $F_{\rm prompt}(x)$ are the $x$ PDFs for non-prompt and prompt $\mathrm{J}/\psi$, respectively while $f^{'}_{\rm B}$ represents the fraction of $\mathrm{J}/\psi$ originating from beauty-hadron decays retrieved from the maximum likelihood fit procedure. The only free parameters in the fitting procedure are $f_{\rm Sig}$ and $f^{'}_{\rm B}$. The latter needs to be corrected for the different acceptance-times-efficiencies for prompt and non-prompt $\mathrm{J}/\psi$, averaged in the $p_{\rm T}$ range where the measurement is performed. These differences can arise from different $\mathrm{J}/\psi$ $p_{\rm T}$ distributions used in the simulations, with different polarization assumptions discussed in the next section. The fraction of non-prompt $\mathrm{J}/\psi$ corrected for these effects, $f_{\rm B}$, is obtained as:

\begin{equation}
\label{eq:fBCorrProcedure} 
f_{\rm B} = \left( 1 + \frac{1-f^{'}_{\rm B}}{f^{'}_{\rm B}} \times \frac{\langle A \times \epsilon \rangle_{\rm B}}{\langle A \times \epsilon \rangle_{\rm prompt}} \right)^{-1},
\end{equation}

where $\langle A \times \epsilon \rangle_{\rm prompt}$ and $\langle A \times \epsilon \rangle_{\rm B}$ represent the average acceptance-times-efficiency values for prompt and non-prompt $\mathrm{J}/\psi$, respectively, in the considered $p_{\rm T}$ interval. 

The various PDFs entering into the determination of $f_{\rm B}$ are described in Refs.~\cite{Abelev:2012gx,Adam:2015rba}. The PDFs corresponding to $\mathrm{J}/\psi$, namely $F_{\rm prompt}(x)$, $F_{\rm B}(x)$ and  $M_{Sig}(m_{\rm ee})$, are determined from Monte Carlo simulations~\cite{ALICE:jpsi:02}. 

\begin{figure}[h!]
  \begin{center}
  \includegraphics[width=0.7\textwidth]{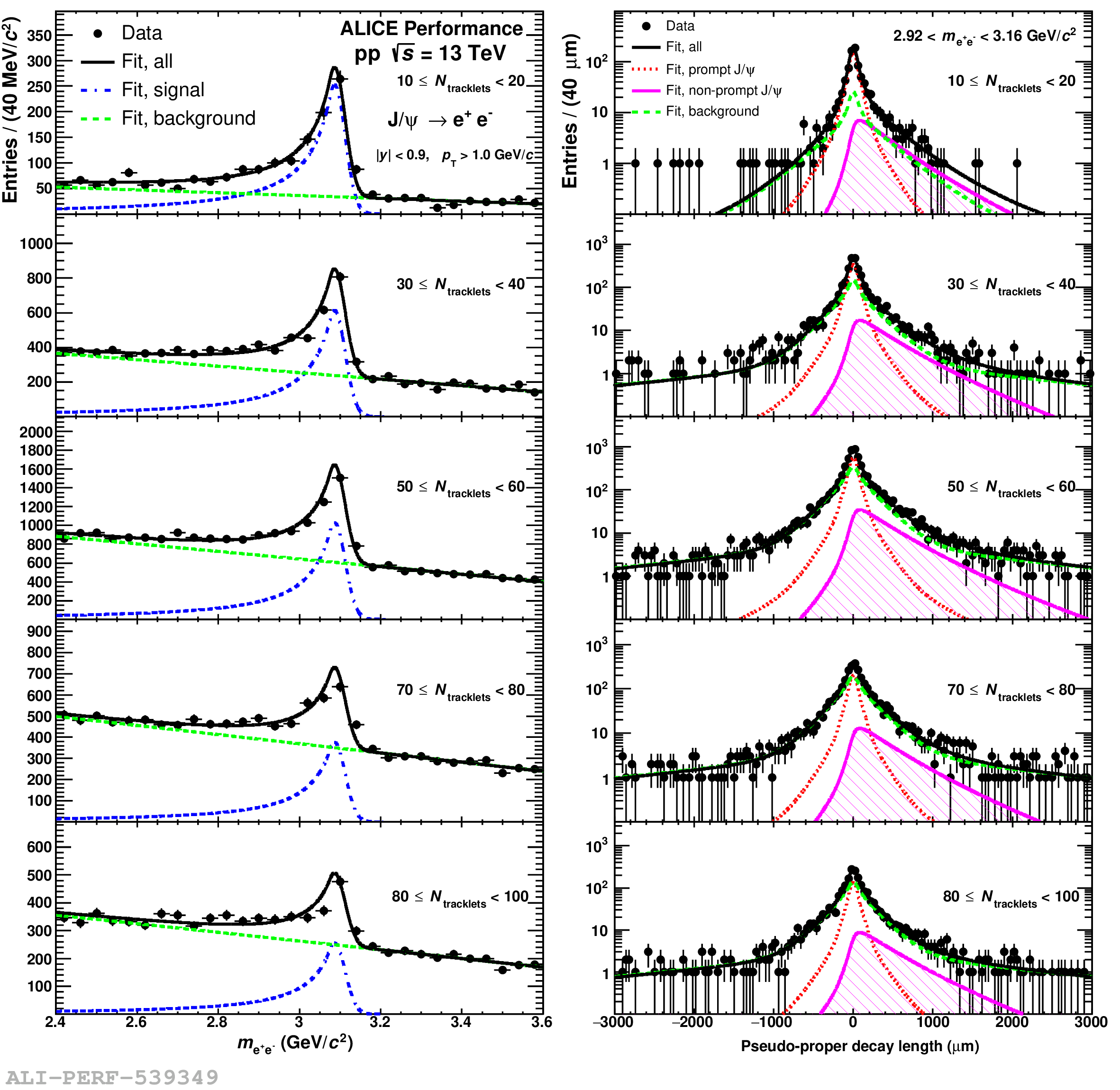}
  \end{center}
  \caption{Invariant mass (left) and pseudoproper decay length (right) distributions for $\mathrm{J}/\psi$ candidates at midrapidity with superimposed projections of the maximum likelihood fit, for $p^{\mathrm{J}/\psi}_{\rm T} > 1~\mathrm{GeV}/c$, performed for different multiplicity intervals with both minimum-bias and high-multiplicity triggered events. This canvas shows the likelihood of fit performance only.}
  \label{Fig:likelihoodFitsMult}
\end{figure}

Examples of dielectron invariant mass and pseudoproper decay length distributions with superimposed projections of the total maximum likelihood fit functions for $p_{\rm T} > 1~\mathrm{GeV}/c$, in multiplicity intervals for the combination of minimum bias and high-multiplicity triggered events are shown in Fig.~\ref{Fig:likelihoodFitsMult}.
\FloatBarrier

%Different components of the likelihood fit function are superimposed on the invariant-mass and pseudoproper decay length distributions of opposite charge-sign candidates. In particular, for the pseudoproper decay length the components relative to the background, prompt and non-prompt $\mathrm{J}/\psi$ are shown. 

\section{Results}
\label{sec:results}

\begin{figure}[!htb]
    \centering
    \includegraphics[width=0.42\columnwidth]{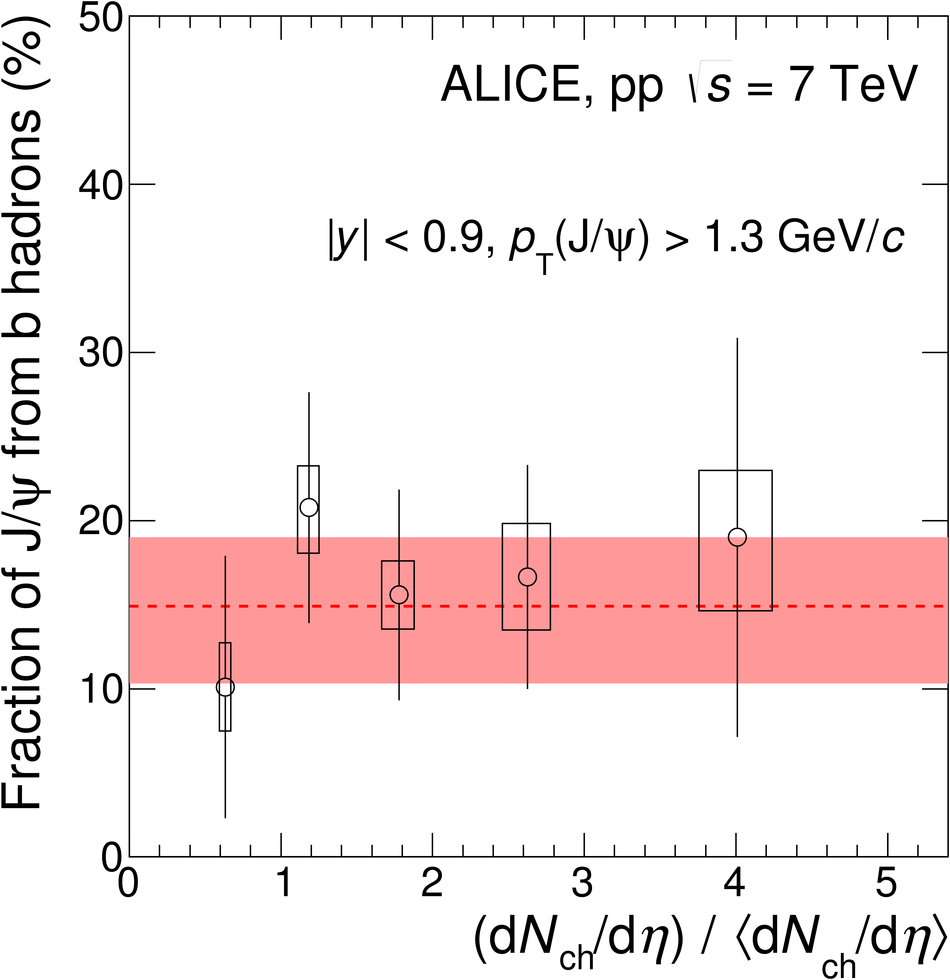}
    \caption{Non-prompt $\mathrm{J}/\psi$ fraction as a function of the relative charged-particle multiplicity at central rapidity for $p_{\rm T} > 1.3~\mathrm{GeV}/c$~\cite{ALICE:jpsi:nonprompt:mult:7tev}. The vertical bars represent the statistical uncertainties, while the empty boxes represent the systematic uncertainties. The width and height of these empty boxes indicate the measurement uncertainty on the horizontal and vertical axes respectively. The dashed line shows the value of $f_{\rm B}$ measured in the same $p_{\rm T}$ range and integrated over multiplicity~\cite{Abelev:2012gx}. The shaded area represents the statistical and systematic uncertainties on the multiplicity-integrated result added in quadrature.}
    \label{fig:fb7TeV}
\end{figure}

%% Non prompt J/psi vs multiplicity
The relative yield of $\mathrm{J}/\psi$ from beauty hadron decays as a function of the charged-particle multiplicity can be evaluated from the inclusive $\mathrm{J}/\psi$ yield and the fraction of non-prompt $\mathrm{J}/\psi$ per multiplicity interval: 
\begin{equation}   
\frac{ \mathrm{d}N_{\mathrm{J}/\psi}^{\rm non-prompt} / \mathrm{d}y }{\langle \mathrm{d}N_{\mathrm{J}/\psi}^{\rm non-prompt} / \mathrm{d}y \rangle }  =
\frac{ {\rm d} N_{\mathrm{J}/\psi} / {\rm d}y }{ \left\langle  {\rm d} N_{\mathrm{J}/\psi} / {\rm d}y \right\rangle }  \cdot
\frac{f_{\rm B}}{\langle f_{\rm B} \rangle}.
\label{eq:JpsiYields}
\end{equation}
where $f_{\rm B}$ is the fraction of non-prompt $\mathrm{J}/\psi$ in each multiplicity interval,
$\langle f_{\rm B} \rangle$ is the fraction in the multiplicity integrated sample~\cite{Abelev:2012gx, ALICE:jpsi:01}, 
and $({\rm d}N_{\mathrm{J}/\psi}/{\rm d}y) \big/ \langle {\rm d}N_{\mathrm{J}/\psi}/{\rm d}y \rangle$ is the inclusive $\mathrm{J}/\psi$ relative yield measured in each multiplicity interval normalized to its value in inelastic pp~collisions~\cite{Abelev:2012rz}. 
All these quantities were measured using the same data sample and the statistical correlations were considered. 
The fraction of non-prompt $\mathrm{J}/\psi$ for $p_{\rm T} > 1.3~\mathrm{GeV}/c$ as a function of the midrapidity multiplicity in pp collisions at $\sqrt{s}$ = 7 TeV is shown in Fig. \ref{fig:fb7TeV}. Similar analysis is being carried out in pp collisions at $\sqrt{s}~=~13~\mathrm{TeV}$.

\begin{figure}[!htbp]
    \centering
    \includegraphics[width=0.4\textwidth]{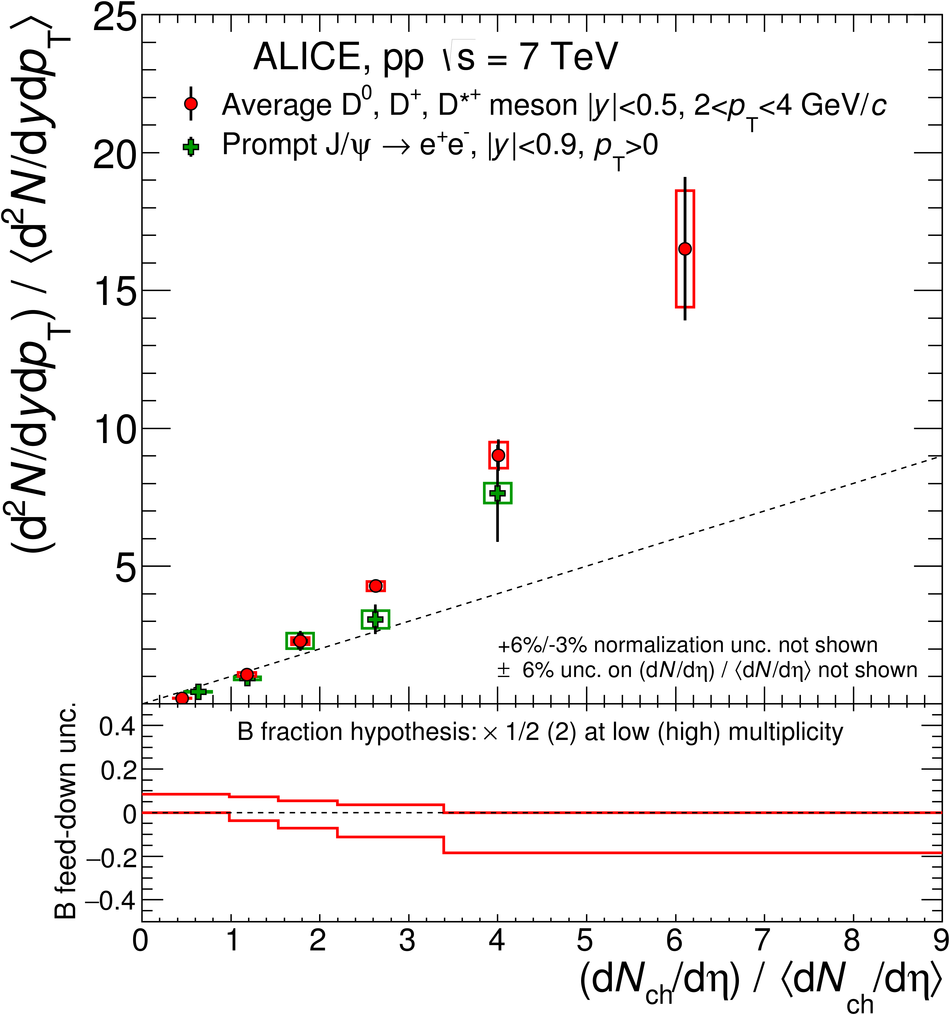}
    \includegraphics[width=0.4\textwidth]{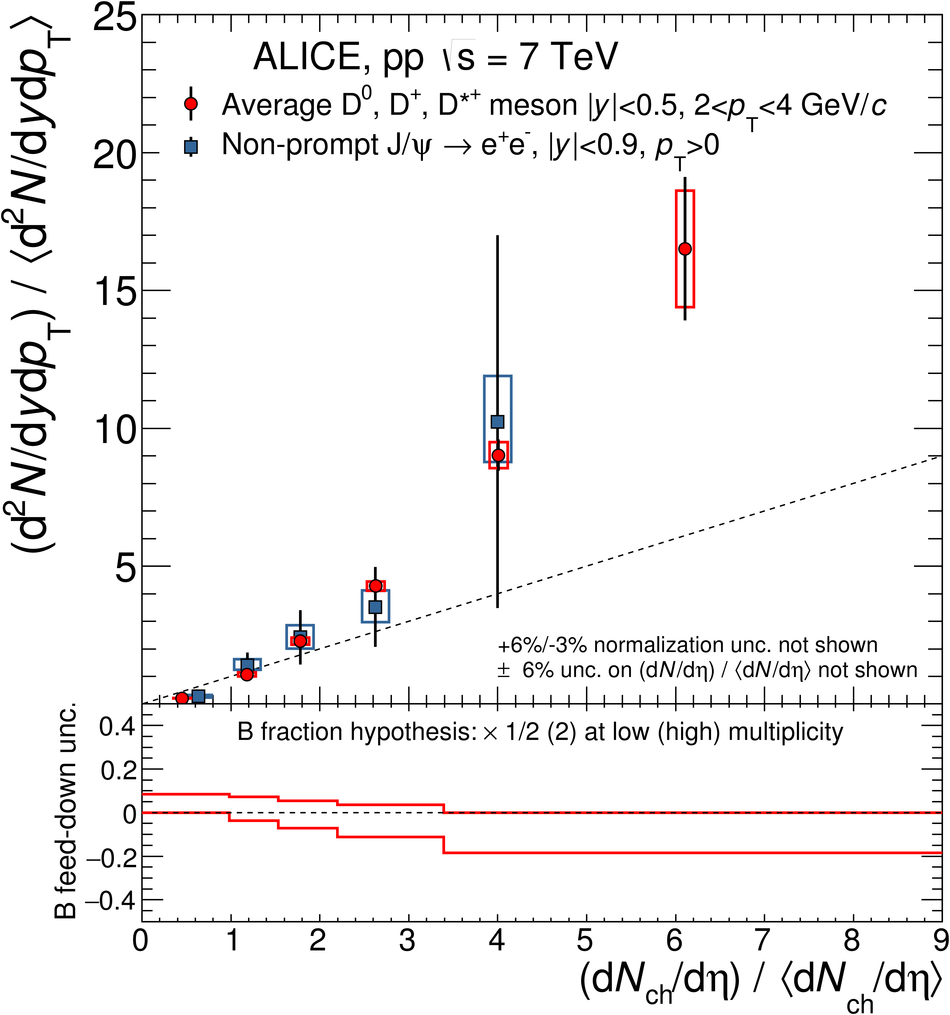}
    \caption{Relative yields of average D meson, prompt (left) and non-prompt (right) $\mathrm{J}/\psi$ as a function of the relative charged-particle multiplicity at midrapidity, for the $p_{\rm T}$-integrated case, in pp collisions at $\sqrt{s}$ = 7 TeV~\cite{ALICE:jpsi:nonprompt:mult:7tev}. The relative yields are presented on the top panels with their statistical (vertical bars) and systematic (boxes) uncertainties, except for the uncertainty on the feed-down fraction for D mesons, which is drawn separately on the bottom panels. The points are located on the x-axis at the average value of $(\mathrm{d}N_{\rm ch}/\mathrm{d}\eta) / \langle \mathrm{d}N_{\rm ch}/\mathrm{d}\eta \rangle$. The dashed line represents a linear trend.}
    \label{fig:jpsiyield7tev}
\end{figure}

Relative yields of prompt and non-prompt $\mathrm{J}/\psi$ calculated with Eq. \ref{eq:JpsiYields} as well as D mesons present a similar increase with midrapidity charged-particle multiplicity (see Fig.~\ref{fig:jpsiyield7tev}). The comparison of open (D meson) and hidden ($\mathrm{J}/\psi$) heavy flavour production suggests that this behaviour is most likely related to the ${\rm c\overline{\rm c}}$ and ${\rm b\overline{\rm b}}$ production processes, and is not significantly influenced by hadronization. The enhancement of the heavy-flavour relative yields with the midrapidity charged-particle multiplicity is qualitatively consistent with the calculations of the contribution from Multiparton Interactions (MPIs) to particle production at LHC energies~\cite{Bartalini:2010su,Sjostrand:1987su,Porteboeuf:2010dw}. 

%%% inclusive Jpsi vs multiplicity 13 tev

\begin{figure}[htb]
    \centering
    \includegraphics[width=0.4\linewidth]{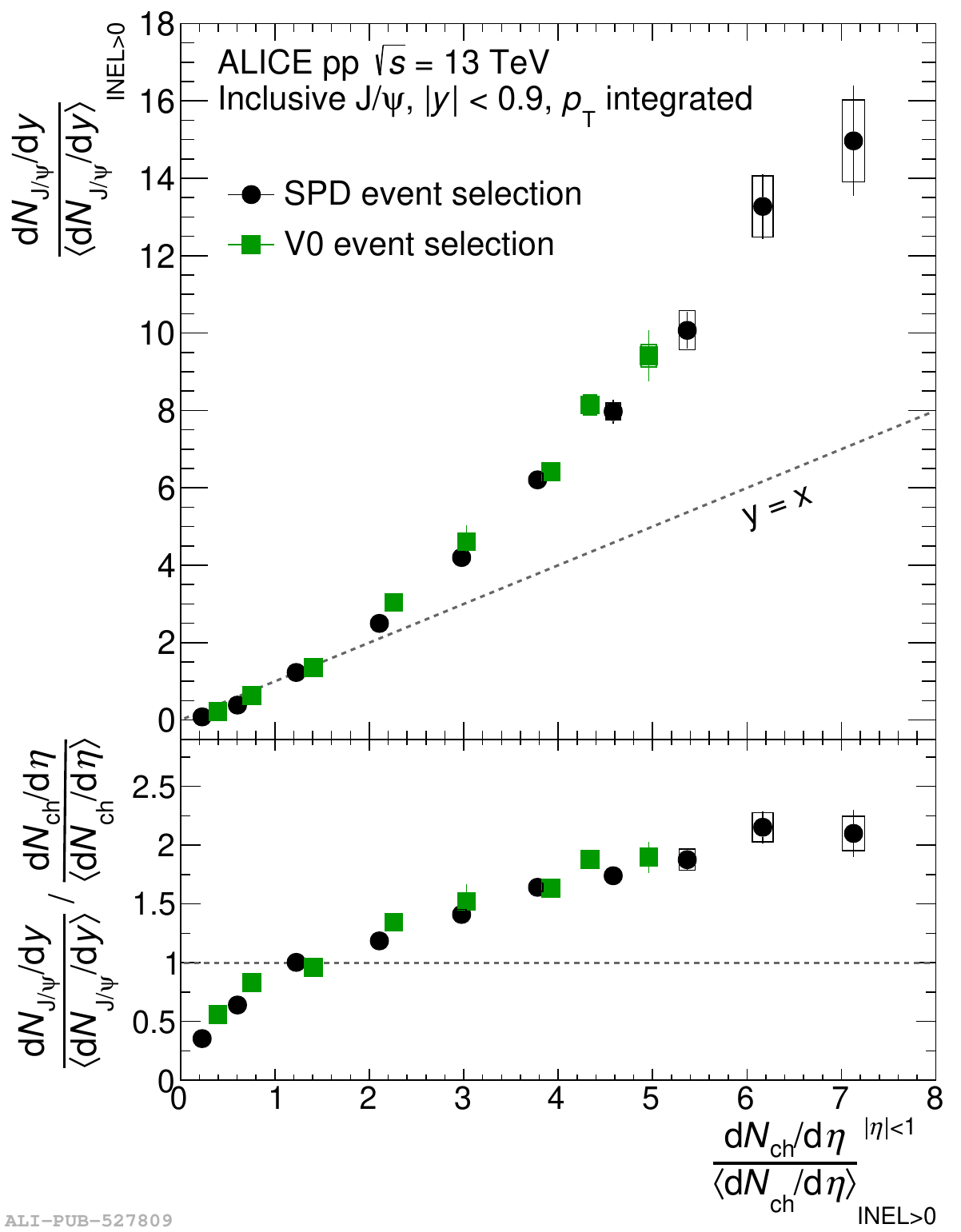}
    \caption{Normalized inclusive $p_{\rm T}$-integrated $\mathrm{J}/\psi$ yield at midrapidity as a function of normalized charged-particle pseudorapidity density at midrapidity ($|\eta|<1$) with the event selection based on SPD tracklets at midrapidity or V0 signals at forward rapidity, in pp collisions at $\sqrt{s}$ = 13 TeV~\cite{ALICE:jpsi:03}. Top: normalized $\mathrm{J}/\psi$ yield (diagonal drawn for reference). Bottom: double ratio of the normalized $\mathrm{J}/\psi$ yield and multiplicity. The error bars show statistical uncertainties and the boxes show the systematic uncertainties.}
    \label{fig_results_mid_fwd}
\end{figure}

The normalized $\mathrm{J}/\psi$ yield as a function of the normalized charged-particle pseudorapidity density at midrapidity, $\mathrm{d}N_{\rm ch}/\mathrm{d}\eta / \langle \mathrm{d}N_{\rm ch}/\mathrm{d}\eta \rangle$, is shown in the top panel of Fig. \ref{fig_results_mid_fwd}, in addition to the double ratio to multiplicity in the bottom panel. The dashed line shown also in the figure is a linear function with a slope of unity. 

\FloatBarrier

The measurements include both the MB and HM triggered events and reach up to 7 times the average charged-particle multiplicity, 
when events are selected based on the measured midrapidity multiplicity. This is a significant extension compared to similar results in pp collisions at $\sqrt{s}$ = 7 TeV~\cite{Abelev:2012rz}, where only the range up to 4 was covered and with larger uncertainties. The results for the two event selection methods by SPD and V0 are in very good agreement.
The normalized $\mathrm{J}/\psi$ yield grows significantly faster than linear with the normalized multiplicity. 
The normalized prompt and non-prompt $\mathrm{J}/\psi$ yield as a function of charged-particle multiplicity in pp collisions at $\sqrt{s}$ = 13 TeV will be calculated by combining the multiplicity-dependent relative yield of inclusive $\mathrm{J}/\psi$ (as shown in Fig. \ref{fig_results_mid_fwd}) and $f_{\rm B}$ (performed by likelihood fit as shown in Fig. \ref{Fig:likelihoodFitsMult}) in a future study. This result could give a new insight into the correlation between heavy-quark production and multiparton interaction in extended multiplicity and energy regions.
\section{Summary}
\label{sec:summary}

The multiplicity dependence of inclusive $\mathrm{J}/\psi$ production and the corresponding fractions of non-prompt $\mathrm{J}/\psi$ in pp collisions at $\sqrt{s}~=~13~\mathrm{TeV}$ are measured or ongoing by ALICE.
A stronger than linear increase of the relative production of inclusive $\mathrm{J}/\psi$ as a function of the midrapidity multiplicity is observed for $p_{\rm T}$-integrated yields, in which the event-multiplicity can reach 7 times the average multiplicity thanks to the high multiplicity triggered events in pp collisions at $\sqrt{s}~=~13~\mathrm{TeV}$.

The measurement of non-prompt $\mathrm{J}/\psi$ as a function of multiplicity at midrapidity has also been studied in pp collisions at $\sqrt{s}$ = 7 TeV with ALICE. The fraction of non-prompt $\mathrm{J}/\psi$ doesn't show a significant dependence on charged-particle multiplicity. A similar increase with the charged-particle multiplicity of normalized prompt and non-prompt $\mathrm{J}/\psi$ yields, as well as D mesons, suggests that heavy-flavour enhancement is not significantly influenced by hadronization but is more likely related to the ${\rm c\overline{c}}$ and ${\rm b\overline{b}}$ production processes. A measurement with higher precision can be done via the data sample collected in pp collisions at $\sqrt{s}$ = 13 TeV by combining the multiplicity-dependent $f_{\rm B}$ and relative yield of inclusive $\mathrm{J}/\psi$ yield measured at the same centre-of-mass energy with the ALICE experiment.

\section*{Acknowledgements}
This work is supported by the National Natural Science Foundation of China (No. 12275103, No. 12061141008).

\bibliographystyle{abbrv}
\bibliography{bibliography}

\end{document}